\documentclass[twocolumn,aps,prb,amsmath,amssymb]{revtex4}

\usepackage{graphicx}
\usepackage{dcolumn}
\usepackage{bm}

\usepackage{hyperref}

\graphicspath{{Figures_spectral_jumps/}} 

\newcommand{\carb}{$^{13}\textrm{C}$ }

\begin{document}

\title{Probing the dynamics of a nuclear spin bath in diamond through time-resolved central spin magnetometry}
\author{A. Dr\'eau$^{1}$}
\author{P. Jamonneau$^{1}$}
\author{O. Gazzano$^{2}$}
\author{S. Kosen$^{1}$}
\author{J.-F. Roch$^{1}$}
\author{J. R. Maze$^{3}$}
\author{V. Jacques$^{1}$}
\affiliation{$^{1}$Laboratoire Aim\'{e} Cotton, CNRS, Universit\'{e} Paris-Sud and Ecole Normale Sup\'erieure de Cachan, 91405 Orsay, France}
\affiliation{$^{2}$ Universit\"at des Saarlandes, Fachrichtung 7.2 (Experimentalphysik), 66123 Saarbr\"ucken, Germany}
\affiliation{$^{3}$Facultad de F\'{i}sica, Pontificia Universidad Cat\'{o}lica de Chile, Santiago 7820436, Chile}

\begin{abstract}
Using fast electron spin resonance spectroscopy of a single nitrogen-vacancy defect in diamond, we demonstrate real-time readout of the Overhauser field produced by its nuclear spin environment under ambient conditions. These measurements enable narrowing the Overhauser field distribution by post-selection, corresponding to a conditional preparation of the nuclear spin bath. Correlations of the Overhauser field fluctuations are quantitatively inferred by analysing the Allan deviation over consecutive measurements. This method allows to extract the dynamics of weakly coupled nuclear spins of the reservoir.

\end{abstract}

\maketitle

\begin{figure}[t]
\begin{centering}
\includegraphics[width=8.5cm]{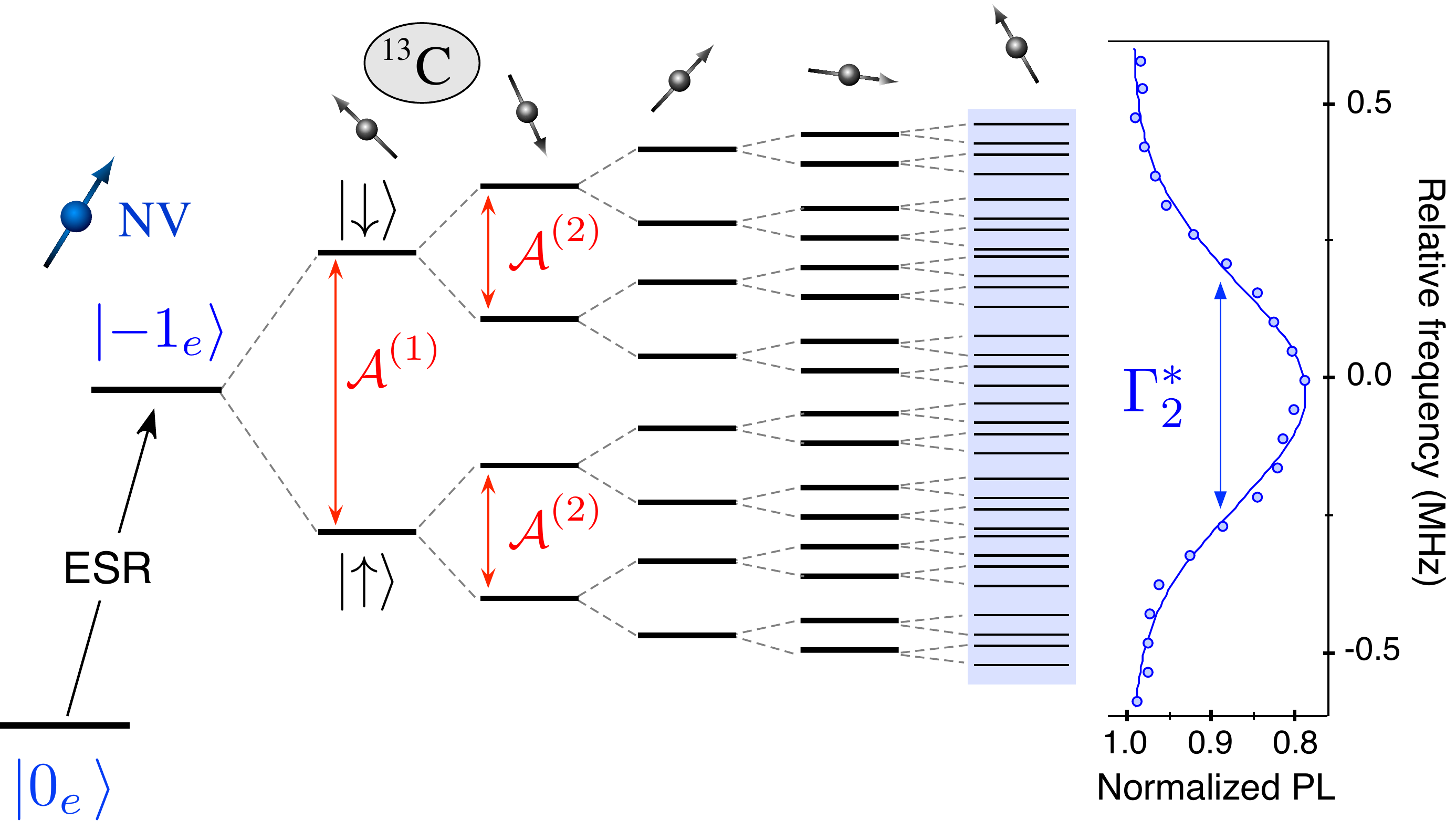}
\caption{(color online) Hyperfine structure of a single NV defect interacting with nearby \carb nuclear spins. Hyperfine sublevels linked to the intrinsic $^{14}$N nuclear spin of the NV defect are not shown. The right panel shows a typical spectrum of the ESR transition between the $m_{s}=0$ [$\left|0_e\right.\rangle$] and $m_{s}=-1$ [$\left|-1_e\right.\rangle$] electron spin sublevels.} 
\label{Fig1}
\end{centering}
\end{figure}

Spins in solids are textbook platforms to model and investigate the dynamics of open quantum systems. An ubiquitous example consists in a central electronic spin, described as a two-level system, interacting through hyperfine coupling with a mesoscopic bath of nuclear spins. This interaction is identified as the major source of decoherence for solid-state spin qubits~\cite{Khaetskii2002,Tyryshkin2012,Gopi2009}. Measuring and controlling the dynamics of such complex environments is therefore a central challenge in quantum physics~\cite{Stepanenko2006,Klauser2006,Bluhm2010,Togan2011}, with potential applications in solid-state quantum information processing~\cite{Ladd_Nature2010} and metrology~\cite{Gary2013,Mamin2013,Staudacher2013}.\\
\indent In this Letter, we explore the dynamics of a dilute nuclear spin bath interacting with the electronic spin of a single nitrogen-vacancy (NV) color center in diamond. This atomic-sized defect has attracted considerable interest over the last years because its ground state is an electronic spin triplet $S=1$ that can be optically initialized, coherently manipulated with microwave magnetic fields and read-out by optical means~\cite{RevueNJP}. In ultrapure diamond samples, decoherence of the NV center electronic spin is mainly caused by interaction with a bath of \carb nuclear spins ($I_c=1/2$) randomly dispersed in the diamond lattice \cite{Gopi2009,Mizuochi2009,Kiu2012,Maze2012}. As shown in Fig.~\ref{Fig1}, each \carb nuclear spin $n$ of the bath induces a hyperfine splitting $\mathcal{A}^{(n)}$ of the NV spin sublevels, whose amplitude depends on its lattice site position with respect to the NV defect~\cite{Gali2008,Gali2009,Smeltzer2011,Dreau2012}. All hyperfine splittings from the nuclear spin bath add up, resulting in a quasi-continuum distribution of hyperfine lines. Each nuclear spin configuration of the bath produces an effective hyperfine magnetic field, colloquially known as ``Overhauser field'', which randomly fluctuates through nuclear spin flips. In most experiments, these fluctuations are much faster than the measurement time scale, so that statistical averaging over all the configurations of the \carb nuclear spin environment leads to an inhomogeneous linewidth $\Gamma_2^*$ of the NV defect electron spin resonance (ESR)~\cite{Kiu2012,Maze2012}, as shown in the right panel of Fig.~\ref{Fig1}. This limitation can be circumvented by performing measurements faster than the correlation time of the nuclear spin bath. This can be achieved at cryogenic temperature by using Overhauser field-selective dark resonances in a $\Lambda$-type level configuration~\cite{Togan2011}. Here we follow an alternative approach which simply consists in acquisitions of optically detected ESR spectra under ambient conditions. The NV defect is used as a magnetometer to infer the instantaneous ``Overhauser field'' and its dynamics through the detection of Zeeman shifts of the ESR frequency induced by nuclear spin flips in the local environment. 
\begin{figure}[t]
\begin{center}
\includegraphics[width=8.65cm]{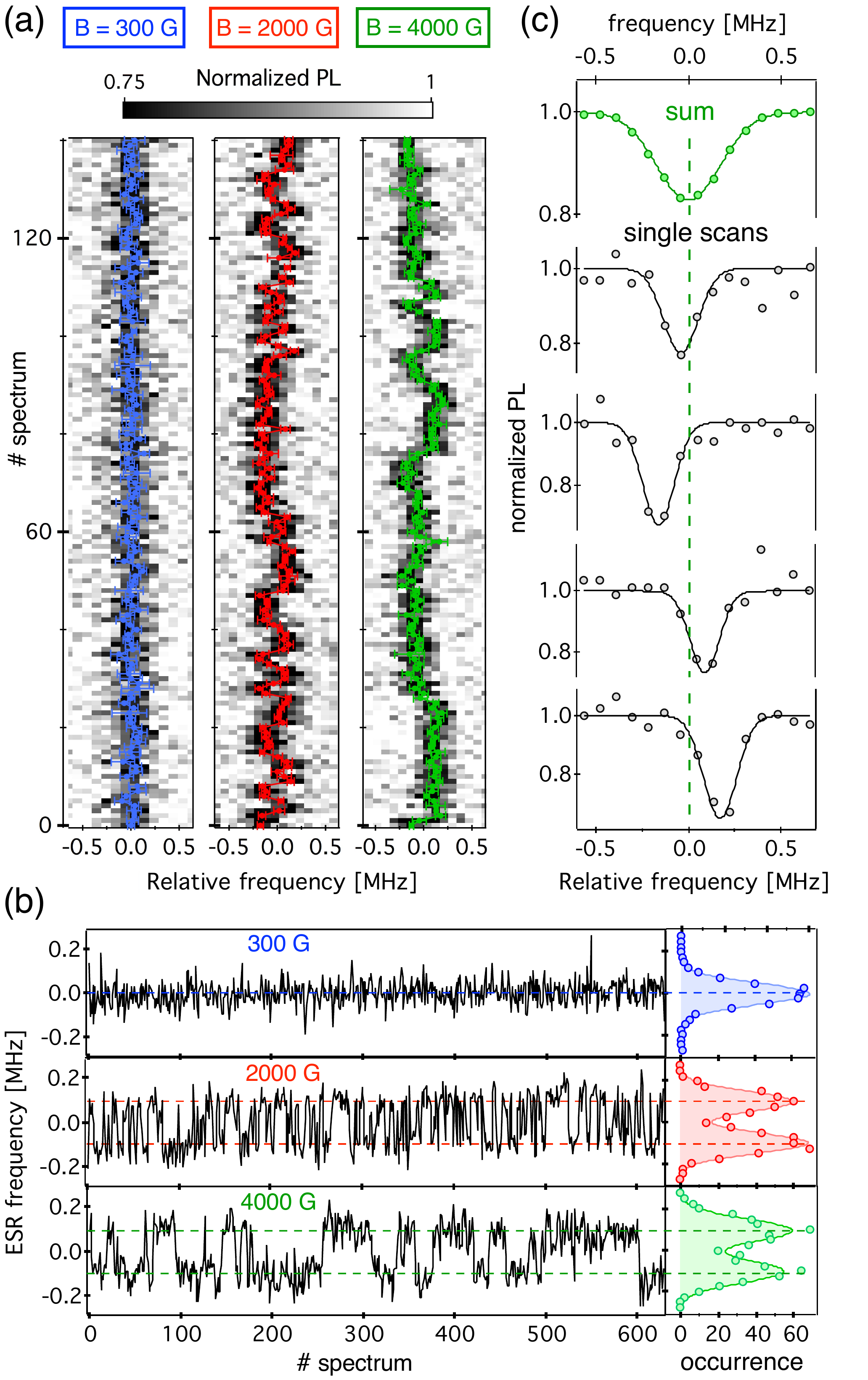}
\caption{(color online) (a) Intensity plots of consecutive ESR spectra recorded at different magnetic field. Pulsed-ESR spectroscopy is performed with a MW $\pi$-pulse duration of 3 $\mu$s. Markers show the ESR frequencies obtained by fitting each individual ESR spectrum with a Gaussian function. The error bar indicate the fit uncertainty with a 95 \% confidence interval. (b) Time evolution of the instantaneous ESR frequency over $630$ consecutive individual spectra recorded at different magnetic fields (total acquisition time $\sim300$~s). The right panels show the corresponding histograms. (c) Top panel: Averaged sum of individual ESR spectra recorded at $B=4000$~G. The inhomogeneous linewidth is $403 \pm 6$~kHz. Lower panels: Selected individual ESR spectra recorded at $B=4000$~G. Statistical analysis over the set of $630$ individual spectra leads to an ESR linewidth of $280\pm70 $~kHz, limited by the $\pi$-pulse duration used for pulsed-ESR spectroscopy.}
\label{Fig2}
\end{center}
\end{figure}

A central idea of this work is to apply a static magnetic field $B$ along the NV defect axis ($z$) in order to tune the correlation time of the nuclear spin environment. Indeed, the flipping rate $\gamma_{R}^{(n)}$ of each \carb of the bath weakly coupled by hyperfine interaction with the NV defect electronic spin can be written as~\cite{Dreau2013} 
\begin{equation}
\gamma_{R}^{(n)}=\frac{1}{T_{R}^{(n)}}\propto  \frac{[\mathcal{A}_{ani}^{(n)}]^2}{[\mathcal{A}_{ani}^{(n)}]^2+(\mathcal{A}_{zz}^{(n)}-\gamma_{n}B)^2} \ ,
\label{T1}
\end{equation}
where $\mathcal{A}_{ani}^{(n)}$ and $\mathcal{A}_{zz}^{(n)}$ are the anisotropic and longitudinal components of the hyperfine tensor, which depends on the lattice site position of the \carb with respect to the NV defect, and $\gamma_{n}\approx 1.07$~kHz/G is the $^{13}$C gyromagnetic ratio. The nuclear spin lifetime $T_{R}^{(n)}$ evolves quadratically with the magnetic field and can reach few seconds at high fields ($B>2000$~G) for \carb with hyperfine coupling strengths $\mathcal{A}^{(n)}$ weaker than $200$~kHz~\cite{Dreau2013,Maurer2012,Waldherr2014}. This is  long enough to be detected through fast ESR spectroscopy under ambient conditions. 

Individual NV defects in a high-purity diamond sample with a natural abundance of \carb ($1.1\%$) are optically addressed at room temperature using a scanning confocal microscope. In such sample, the inhomogeneous ESR linewidth $\Gamma_2^*$ is few hundreds kHz [Fig.~\ref{Fig1}]~\cite{Mizuochi2009,Kiu2012,Maze2012}. A permanent magnet placed on a three-axis translation stage is used to apply a static magnetic field along the NV defect axis ($z$). Spectroscopy of the ESR transition between the $m_{s}=0$ and $m_{s}=-1$ electron spin sublevels is performed through repetitive excitation of the NV defect with a resonant microwave (MW) $\pi$-pulse followed by a $300$-ns read-out laser pulse~\cite{Dreau2011}. This sequence is continuously repeated during 30 ms while recording the spin-dependent photoluminescence (PL) intensity of the NV defect, before incrementing the MW frequency. All ESR spectra shown in this work were recorded on 15 points, corresponding to a measurement time $T_m \sim 450$~ms per spectrum. In practice, three MW sources are swept simultaneously in order to optimize the ESR contrast and to get rid of the dynamics linked to the intrinsic $^{14}$N nuclear spin ($I_N=1$) of the NV defect~\cite{Neumann_Science2010}. The MW sources are synchronized and their frequencies detuned by $\mathcal{A}_N=2.16$~MHz, which correspond to the hyperfine splitting induced by the $^{14}$N nucleus~\cite{sup}.

\begin{figure*}[t]
\begin{centering}
\includegraphics[width=17cm]{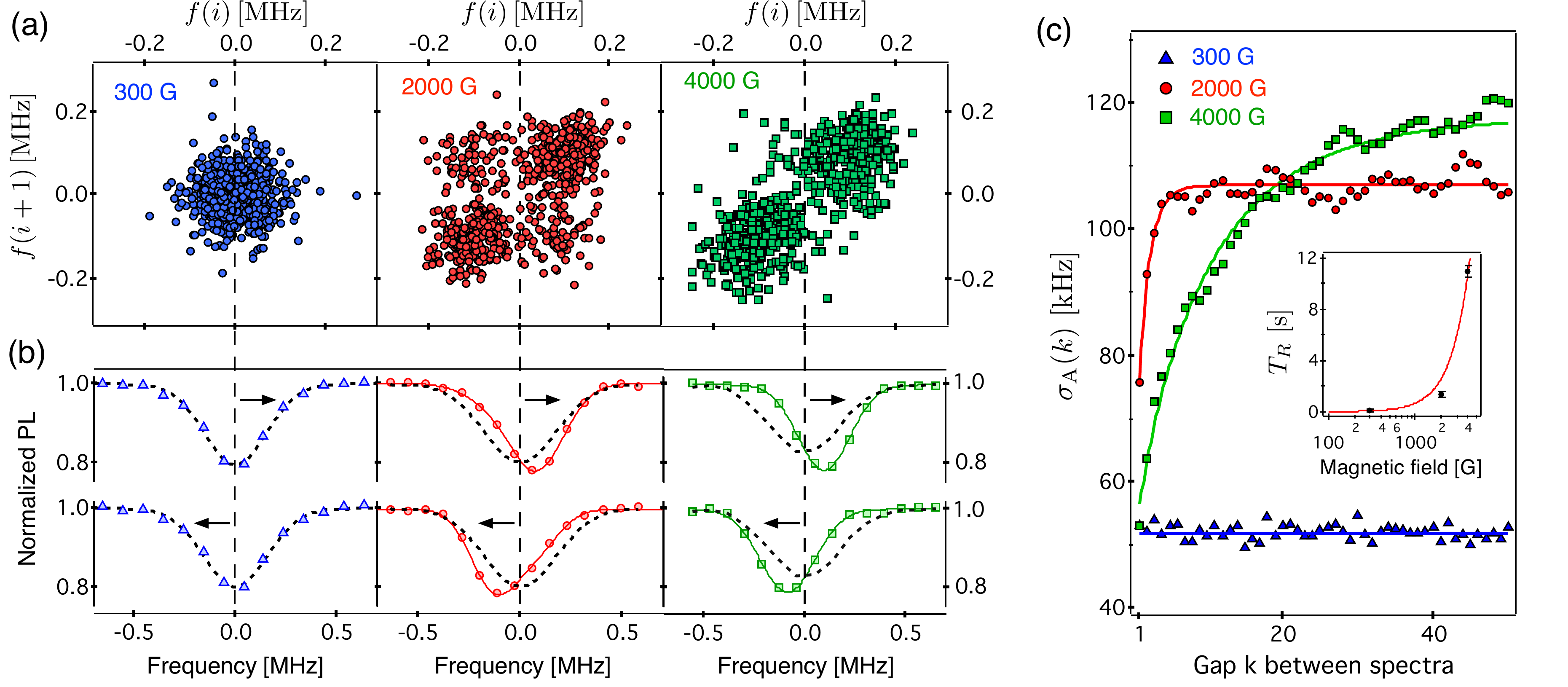}
\caption{(color online) (a) Distribution of consecutive ESR frequencies at three different magnetic fields. (b) Post-selected ESR spectra obtained by summing experimental runs $i+1$ conditioned by $f(i)>0$ (upper panels) or $f(i)<0$ (lower panels). The black dashed line correspond to ESR spectra averaged over all individual runs. Solid lines are fits with Gaussian functions. At $B=4000$~G, the linewidth of the post-selected ESR spectrum is reduced to $303\pm 5$~kHz
. (c) Allan deviation $\sigma_{\rm A}(k)$ inferred from the set of ESR frequencies $\{f(i)\}$ at different magnetic fields. Solid lines are data fitting with Eq. (\ref{eq_Allan_dev_model}) leading to $\mathcal{A}^{(1)}=210\pm7$~kHz. Inset: Relaxation time $T_R$ inferred from data fitting as a function of the magnetic field. The solid line is a quadratic fit as predicted by Eq.~(\ref{T1}).}
\label{Fig3}
\end{centering}
\end{figure*}

Intensity plots of consecutive ESR spectra recorded from a single NV defect are depicted on Fig.~\ref{Fig2}(a) for three different magnetic field amplitudes. For each individual spectrum $i$, the instantaneous ESR frequency $f(i)$ was extracted through data fitting with a Gaussian function. The time evolution of $f(i)$, which mirrors the one of the Overhauser field, is shown in Fig.~\ref{Fig2}(b). While the ESR frequency does not exhibit significant fluctuations in time at low field ($B=300$~G), well-resolved spectral jumps can be observed when the magnetic field is increased. In this high magnetic field regime, the correlation time of the bath becomes longer than the measurement time $T_m$, which enables measuring the instantaneous Overhauser field produced by different configurations of the nuclear spin bath. This is further illustrated in Fig.~\ref{Fig2}(c) where individual ESR spectra recorded at $B=4000$~G are plotted together with the averaged sum of experimental scans (top panel). The instantaneous ESR frequencies of individual runs evolve in time and their linewidths are smaller than the one obtained by averaging over all the configurations of the bath. As expected, spectral narrowing is also accompanied by an improved ESR contrast.\\
\indent In these experiments, the amplitude of the Overhauser field fluctuations is dominated by the dynamics of the nearest \carb nuclear spin of the bath, which induces a hyperfine splitting of the ESR line $\mathcal{A}^{(1)}\approx \mathcal{A}_{zz}^{(1)}\approx 200$~kHz [cf. Fig.~\ref{Fig1}]~\cite{Note}. This hyperfine structure is revealed by the histograms of the instantaneous ESR frequencies $f(i)$ shown in Fig.~\ref{Fig2}(b). At low field the histogram is well described by a Normal distribution, while two peaks separated by 201 $\pm$ 3 kHz can be observed at high field. For this particular $^{13}$C, the anisotropic component of the hyperfine tensor is weak, leading to a relaxation time exceeding seconds at high magnetic fields~\cite{Dreau2013}. We note that the two peaks of the distribution also get broader at high magnetic field, which qualitatively indicates a contribution to the Overhauser field fluctuations from other \carb of the reservoir~\cite{sup}. \\
\indent Measurements of the instantaneous ESR frequency can be used for narrowing down the Overhauser field distribution by post-selection. For that purpose, we first consider the correlations between consecutive measurements, $f(i)$ and $f(i+1)$, whose conditional distributions are shown on Fig.~\ref{Fig3}(a) for different magnetic field amplitudes. From this set of measurements, post-selection of individual ESR spectra is performed depending on the sign of the instantaneous ESR frequency. More precisely, for all measurements that satisfy $f(i)>0$ [or $f(i)<0$], we extract a post-selected ESR spectrum by summing the set of individual runs $i+1$. The results are shown in Fig.~\ref{Fig3}(b). At low magnetic field, post-selected ESR spectra are identical to the one obtained by averaging over all individual runs [black dashed line in Fig.~\ref{Fig3}(b)], pointing out the absence of correlation between consecutive measurements. At higher fields, post-selected ESR spectra have a narrower linewidth ($\sim300$~kHz) with a shifted central frequency, corresponding to the hyperfine splitting of the nearest \carb nuclear spin of the bath. Here conditional measurements enable narrowing down the Overhauser field distribution by polarizing this particular \carb of the bath, in a same way as in recent single shot readout experiments~\cite{Dreau2013,Maurer2012,Comment}.\\
\indent The linewidth of post-selected ESR spectra results from the convolution of the MW excitation spectral profile and the Overhauser field distribution within the acquisition time. For a $\pi$-pulse duration of $3$~$\mu$s, the width of the MW spectral profile is $\sim300$~kHz, which is the main limitation to the ESR linewidth. Although increasing the $\pi$-pulse duration could in principle lead to further spectral narrowing, another consequence would be an overall reduction of the PL signal, which is limited by the duty cycle of the laser pulses in the ESR sequence. In order to keep a high signal-to-noise ratio, the measurement time per spectrum would then need to be increased, which results in averaging over a broader Overhauser field distribution. 
 \\
\begin{figure}[t]
\includegraphics[width=8.7cm]{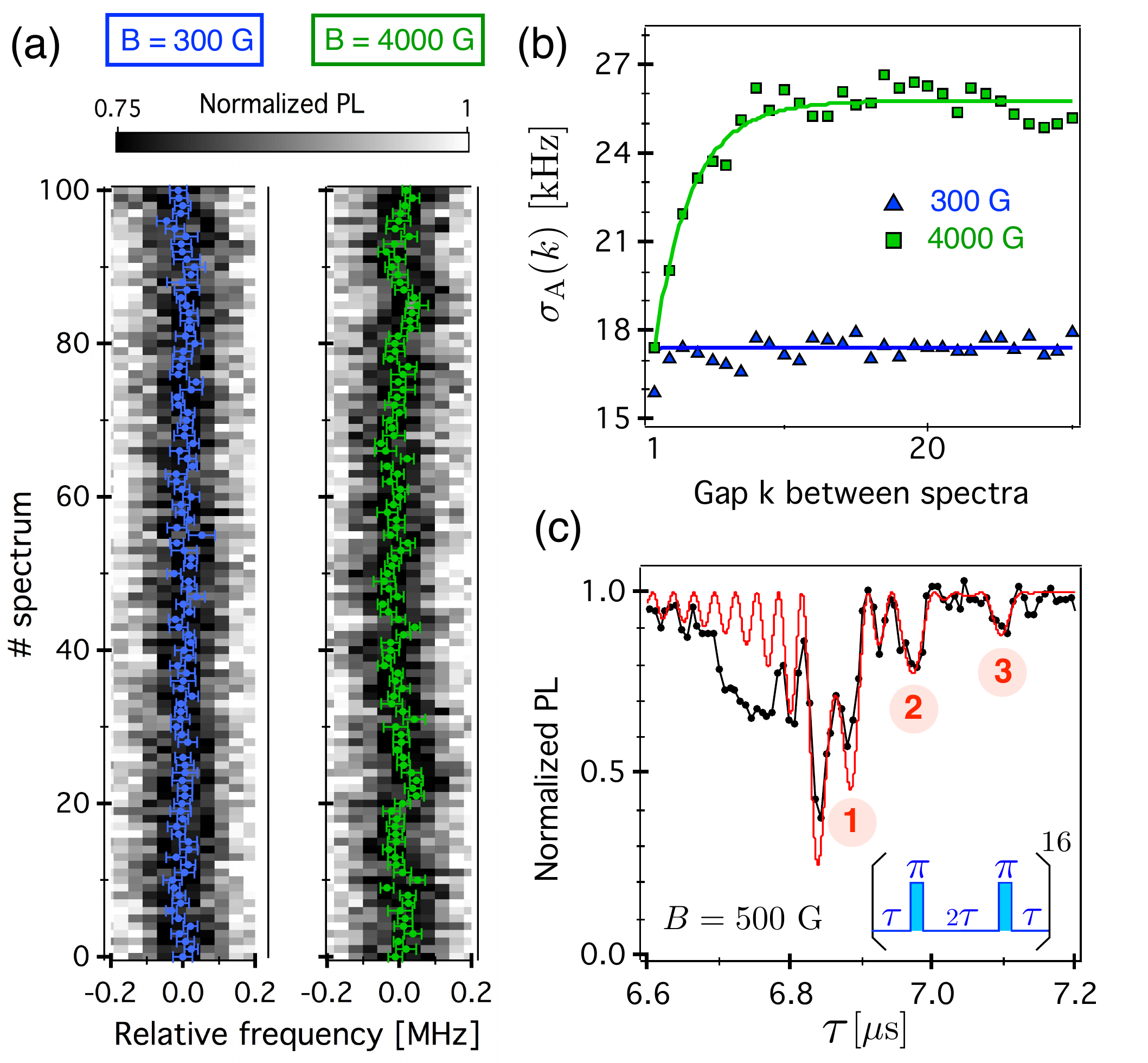}
\caption{(color online) (a) Intensity plots of consecutive ESR spectra recorded for NV2  with a MW $\pi$-pulse duration of 4 $\mu$s. (b) Allan deviations inferred from a set of $300$ consecutive ESR spectra at different magnetic fields. Solid lines are data fitting with Eq.(\ref{eq_Allan_dev_model}), as explained in the main text. At high field, we obtain $T_R=2.6 \pm 0.2$~s and $\sqrt{[\mathcal{A}_{zz}^{(2)}]^2+[\mathcal{A}_{zz}^{(3)}]^2}= 45 \pm 2 $ kHz. The latter value is in decent agreement with the one inferred from the dynamical decoupling signal [see Tab.~\ref{Table1}].(c) Dynamical-decoupling signal obtained by implementing a 32-pulse sequence (see inset) with a magnetic field $B=500$~G. The solid line is data fitting using the procedure described in Ref.\cite{Taminiau2012}. More details can be found in~\cite{sup}.}
\label{Fig4}
\end{figure}
\indent We now analyze more quantitatively the correlations between distant individual ESR spectra by using the Allan deviation $\sigma_{\rm A}$. This statistical tool is commonly used for quantifying the stability of oscillators~\cite{Allan} and is defined as 
	\begin{equation}
	\sigma_{\rm A}(k) = \sqrt{\frac{1}{2}\left\langle [f(i)-f(i+k)]^2 \right \rangle_{\{ i \}} } \, ,
	\label{eq_Allan_dev}
	\end{equation}

where $\langle ... \rangle_{\{ i \}}$ denotes the average over all the data set $\{f(i)\}$. As shown in Fig. \ref{Fig3}(c), the Allan deviation is flat at low field while increasing the separation $k$ between the measurements, as expected for a regime without correlation. At high fields, the Allan deviation first increases with $k$ before reaching a plateau. This behavior indicates strong correlations over distant individual runs. To infer quantitative informations, we derive an analytical expression of the Allan deviation~\cite{sup}
\begin{equation}
\sigma_{\rm A}(k) = \sqrt{\sum_n\frac{{\mathcal{A}_{zz}^{(n)}}^2}{4} \left[ \alpha^{(n)}-{\beta^{(n)}}^2 e^{-2(k-1)T_m)/T_R^{(n)}} \right]+ \sigma_{\rm sn}^2} \, ,
\label{eq_Allan_dev_model}
\end{equation}
with $$\alpha^{(n)} = \frac{T_R^{(n)}}{T_m} (1- \beta)  \,Ê\, {\rm and} \, \, \, \beta^{(n)} = \frac{1- e^{-2 T_m/T_R^{(n)}}}{2 T_m/T_R^{(n)}} \ .$$ 
Here the sum runs over all \carb nuclear spin $n$ of the bath and $\sigma_{\rm sn}$ is the standard deviation of the measurement noise, mostly induced by shot-noise in the detection of the NV center PL. This formula is simplified by considering that correlations are dominated by hyperfine interaction with the nearest \carb ~\cite{sup} and used to fit the experimental data [solid lines in Fig. \ref{Fig3}(c)]. We obtain $\mathcal{A}_{zz}^{(1)}=210\pm7$~kHz in good agreement with the value obtained by other methods [see Fig. \ref{Fig2}(b)]. As expected, the correlation time $T_R^{(1)}$ inferred from the fit increases with the magnetic field [see inset in Fig. \ref{Fig3}(c)], reaching more than ten seconds at $4000$~G. \\
\begin{table}[t]
\caption{\label{Table1} Longitudinal $\mathcal{A}_{zz}^{(n)}$ and anisotropic $\mathcal{A}_{ani}^{(n)}$ hyperfine components of the three \carb nuclear spins detected in the dynamical decoupling signal shown in Fig. \ref{Fig4}(c).}
\begin{ruledtabular}
\begin{tabular}{ccc}
 \carb number $n$ &$\mathcal{A}_{zz}^{(n)}$~[kHz]&$\mathcal{A}_{ani}^{(n)}$~[kHz]\\
\hline
	 1 & -27 $\pm$ 3 & 128 $\pm$ 2  \\ 
    		2 & -28 $\pm$ 2  &  19 $\pm$ 3 \\ 
    		3 & -46 $\pm$ 2 &  20 $\pm$ 3   \\ 
   \end{tabular}
\end{ruledtabular}
\end{table}
\indent Every single NV defect has a specific nuclear spin environment since \carb are randomly placed in the diamond lattice. The fluctuations of the Overhauser field are therefore expected to be different for each NV defect. To illustrate this point, the experiments were repeated with a single NV defect (denoted NV2) for which the amplitude of the Overhauser field fluctuations is not dominated by the nearest \carb nuclear spin. Here the fluctuations of the instantaneous ESR frequency can hardly be observed in consecutive ESR spectra [Fig.~\ref{Fig4}(a)]. However, the Allan deviation indicates unambiguously correlations between individual runs at high magnetic fields [Fig.~\ref{Fig4}(b)]. This observation results from the slowdown in the dynamics of some \carb nuclear spins surrounding the NV defect. To check this assumption, the local \carb environment  was investigated by implementing a 32-pulse dynamical decoupling sequence~\cite{sup}, which enables characterizing individual \carb with hyperfine coupling strengths much smaller than the inhomogeneous ESR linewidth~\cite{Taminiau2012,Kolkowitz2012,Zhao2012}. As shown in Fig.~\ref{Fig4}(c), sharp dips in the signal reveal hyperfine coupling with three individual \carb nuclear spins. Table \ref{Table1} summarizes the values of the longitudinal and anisotropic components of the hyperfine tensor extracted from the dynamical decoupling signal, following the procedure described in Ref.~\cite{Taminiau2012}. For this NV defect, it is very unlikely to observe the dynamics linked to the most strongly coupled \carb owing to the high value of its anisotropic hyperfine component. On the other hand, this component is much smaller ($\simeq 20$ kHz) for the two other \carb nuclear spins leading to long correlation times at high magnetic fields [see Eq.~(\ref{T1})]. Fitting the allan deviation with Eq.~(\ref{eq_Allan_dev_model}) while considering identical relaxation time for the two \carb leads to  $T_R=2.6 \pm 0.2$~s at $B=4000$~G. These experiments demonstrate how the Allan deviation can be used to infer the relaxation time of \carb with coupling strengths that are one order of magnitude smaller than the inhomogeneous dephasing rate of the NV defect. 

In conclusion, we have used a single NV defect in diamond as a highly sensitive magnetometer to measure in real-time the Overhauser field produced by its nuclear spin environment under ambient conditions. Analysis of the Overhauser field fluctuations was achieved by implementing a correlation detection method based on the Allan deviation that extracts the dynamics of weakly coupled nuclear spins of the reservoir. In addition, we have reported narrowing of the Overhauser field distribution through conditional preparation of the nuclear spin bath by post-selection. Further improvements could be achieved by using stronger magnetic fields and/or by decreasing the measurement time, either by improving the collection efficiency with diamond photonic nanostructures~\cite{Babinec_NatNano2010,Neu2014} or by performing single-shot readout of the NV electron spin under a cryogenic environment~\cite{Robledo2011}. These methods might find applications in the context of quantum feedback control and metrology. \\

We thank F.~Grosshans, J. P. Tetienne and G. H\'etet for fruitful discussions. This work is supported by the French National Research Agency (ANR) through the projects A{\sc dvice} and Q{\sc invc}, and by the European Community's Seventh Framework Programme (FP7/2007-2013) under Grant Agreement No. 611143 (D{\sc iadems}). J.R.M acknowledges support from Conicyt grants Fondecyt No. 1141185, PIA programs ACT1108 and ACT1112, and Millennium Scientific Initiative P10-035-F.

\section*{SUPPLEMENTARY INFORMATION}
\subsection{Experimental methods}

We study native NV defects hosted in a commercial [100]-oriented high-purity diamond crystal grown by chemical vapor deposition (Element6) with a natural abundance of ${^{13}}$C isotopes ($1.1\%$). 
Individual NV defects are optically isolated at room temperature using a home-built scanning confocal microscope under optical excitation at $532$~nm. Coherent manipulation of the NV defect electron spin is achieved by applying a microwave field through a copper microwire directly spanned on the diamond surface. Details about the experimental setup can be found in Ref.~[\onlinecite{Dreau2011}]

\begin{figure}[h!]
\begin{centering}
\includegraphics[width=8.7cm]{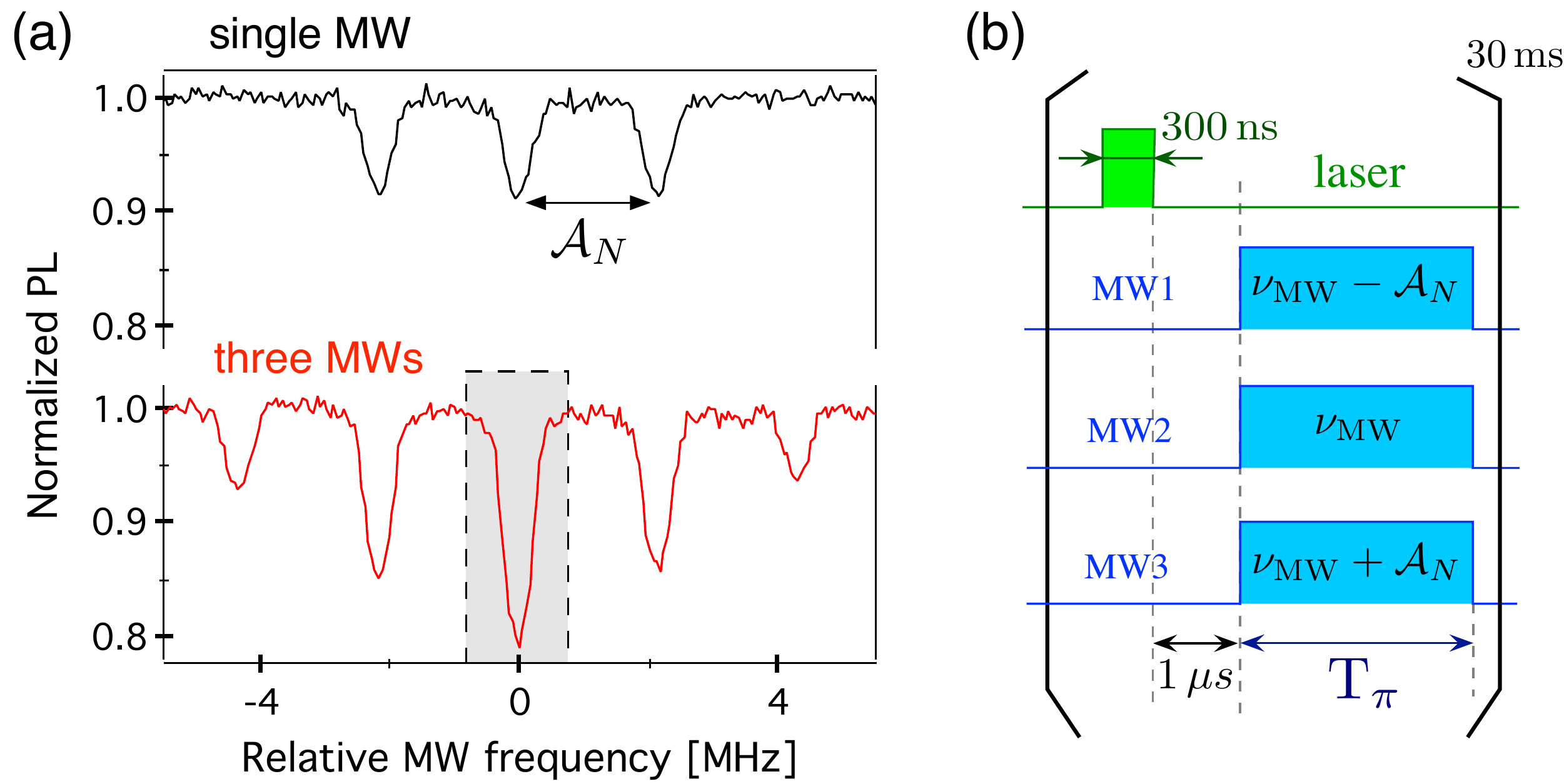}
\caption{(a) ESR spectra recorded by sweeping a single MW frequency (top trace) and three MW frequencies (bottom trace) detuned by $\mathcal{A}_N=2.16$~MHz, which correspond to the hyperfine splitting induced by the $^{14}$N nuclear spin. (b) Experimental pulse sequence used to record the ESR spectra shown in the bottom trace in (a). The sequence is repeated during 30 ms before incrementing the MW frequency $\nu_{\rm MW}$. The dashed rectangle in (a) indicates the frequency range on which the ESR spectra shown in the main text are recorded.}
\label{Fig1_SI}
\end{centering}
\end{figure}

As indicated in the main text, electron spin resonance (ESR) spectroscopy is performed through repetitive excitation of the NV defect with a resonant microwave $\pi$-pulse followed by a $300$-ns read-out laser pulse~[\onlinecite{Dreau2011}]. ESR spectra are recorded by continuously repeating this sequence while sweeping the $\pi$-pulse frequency and recording the spin-dependent PL intensity. A typical spectrum of the ESR transition between the $m_{s}=0$ and $m_{s}=-1$ electron spin sublevels is shown in Fig.~\ref{Fig1_SI}(a) [top trace], revealing the characteristic hyperfine splitting $\mathcal{A}_N=2.16$~MHz linked to the intrinsic $^{14}$N nuclear spin of the NV defect. In order to achieve a maximum contrast of the ESR line and to eliminate the dynamics of the $^{14}$N nuclear spin, three synchronized MW sources were swept simultaneously with a frequency detuning set to $\mathcal{A}_N=2.16$~MHz [Fig.~\ref{Fig1_SI}(b)]. Using such a procedure, the ESR spectrum exhibits \emph{artificially} five resonance lines [Fig.~\ref{Fig1_SI}(a), bottom trace]. The contrast of the central line is maximal because in that case all hyperfine sublevels linked to the $^{14}$N nuclear spin are simultaneously excited. All ESR spectra shown in the main text correspond to a zoom around the central resonance line. The dead-time between consecutive measurements is $30$ ms.

\subsection{Allan deviation modeling}

In the main article, the study is focused on the transition between the $m_{s}=0$ and $m_{s}=-1$ electron spin sublevels of the NV defect. The corresponding ESR frequency $f_{\rm av}$ measured by averaging over all the configurations of the nuclear spin environment is given by $f_{\rm av}=D-\gamma_eB$, where $D$ is the zero-field splitting, $\gamma_e\approx 2.8$~MHz.G$^{-1}$ is the electron gyromagnetic ratio and $B$ is the amplitude of the magnetic field applied along the NV defect quantization axis. Using $f_{\rm av}$ as a frequency reference, we infer the relative frequency shift denoted $f(i)$ by fitting each individual ESR spectrum $i$ with a Gaussian function.

In this section, we derive an analytical formula of the Allan deviation $\sigma_{\rm A}(k)$ defined by
	\begin{equation}
	\sigma_{\rm A}(k) = \sqrt{\frac{1}{2}\left\langle [f(i)-f(i+k)]^2 \right \rangle_{\{ i \}} } \, ,
	\label{eq_Allan_dev}
	\end{equation}
where $\langle ... \rangle_{\{ i \}}$ denotes the average over all the data set $\{f(i)\}$. In the following, we rather use the temporal variables $t$ and $\tau$ which denote the time at which the ESR spectrum is recorded and the delay between ESR spectra $i$ and $i+k$, respectively. Neglecting the dead time between consecutive measurements, we have $\tau = k \cdot T_m$, where $T_m \simeq 0.45$ s is the time needed to acquire an individual ESR spectrum. Using these variables, the Allan deviation writes
	\begin{subequations}
	\begin{align}
	\sigma_{\rm A}(\tau)& = \sqrt{\frac{1}{2} \left\langle (f(t+\tau) - f(t))^2 \right \rangle} \\
		& =  \left\{ \frac{1}{2} \left( \left \langle f(t+ \tau) ^2 \right \rangle +  \left \langle f(t)^2 \right \rangle - 2 \left \langle f(t+ \tau) f(t) \right \rangle \right)\right\}^{1/2}\, .
	\label{eq_Allan_dev_temp}
	\end{align}
	\label{test}
	\end{subequations}
where $f(t)$ is a random variable corresponding to the ESR frequencies measured at time $t$.

\subsubsection{Allan deviation considering instantaneous measurements}

We first consider an instantaneous and noiseless ESR measurement for a NV defect coupled with only one \carb nuclear spin ($I=1/2$), leading to a hyperfine splitting $\mathcal{A}$. The relative ESR frequency shift induced by nuclear spin flips can thus take only two values, $+\mathcal{A}/2$ or $-\mathcal{A}/2$, and can be written as
	\begin{equation}
	f(t)= f(0) (-1)^{N(t)} \, .
	\label{eq_f_instantaneous}
	\end{equation}
Here $f(0)=\pm\mathcal{A}/2$ and $N(t)$ is a discrete random variable describing the number of nuclear spin flips occurring within time $t$, whose probability distribution follows a Poisson's law 
	\begin{equation}
	\mathcal{P}[N(t)=j] = \frac{1}{j ! } \left(\frac{t}{T_R}\right)^j e^{-t/T_R} \, ,
	\label{Poisson}
	\end{equation}
where $T_R$ is the \carb nuclear spin relaxation time. The average value of the ESR frequency measured at time $t$ is therefore given by  
	\begin{equation}
	\langle f(t)\rangle = f(0) \sum_j \left(\frac{t}{T_R}\right)^j \frac{(-1)^{j}}{j !}e^{-t/T_R} = f(0) e^{-2 t/T_R} \, .
	\label{eq_f_inst_mean}
	\end{equation}
	
The correlation function can be expressed in the same way 
	\begin{equation}
	\langle f(t) f(t+\tau) \rangle  = f(0)^2 \langle (-1)^{N(t)+N(t+ \tau)} \rangle \, .
	\end{equation}
Assuming that $\tau> 0$, we can write $N(t+ \tau) = N(t) + N(\tau)$. Here the two random variables $N(t)$ and $N(\tau)$ are independent and both follow a Poisson distribution. The correlation function then writes 
	\begin{equation}
	\langle f(t) f(t+\tau) \rangle = f(0)^2 \langle (-1)^{2N(t)}\rangle  \langle (-1)^{N(\tau)} \rangle  =f(0)^2 e^{-2 \tau /T_R} \, .
	\label{eq_f_corr}
	\end{equation}

Using Eq. (\ref{eq_Allan_dev_temp}), the Allan deviation linked to instantaneous ESR frequency measurements for a NV defect coupled with only one \carb nuclear spin is given by
	\begin{equation}
	\sigma_{\rm A}(\tau) = \frac{\mathcal{A}}{2} \sqrt{1-e^{-2\tau/T_R}} \, .
	\end{equation}

This last expression is not realistic since the ESR frequency measurement is prone to errors, due to the combination of a limited ESR contrast, a non-zero linewidth and shot noise in the detection of the NV defect PL intensity. We therefore include an overall measurement noise by expressing the measured ESR frequency as 
	\begin{equation}
	f(t) = f(0)(-1)^{N(t)} + \tilde{\sigma}_{\rm sn} (t) \, ,
	\end{equation}
where $\tilde{\sigma}_{\rm sn} (t)$ is a random variable associated to the measurement noise with mean value equals to zero and standard deviation $\sigma_{\rm sn}$. We assume that $\tilde{\sigma}_{\rm sn} (t)$ and $\tilde{\sigma}_{\rm sn} (t+\tau)$ are independent variables, without correlations with $N(t)$. The Allan deviation then writes
	\begin{equation}
	\sigma_{\rm A}(\tau) = \sqrt{\frac{{\mathcal{A}}^2}{4}  (1-e^{-2\tau/T_R})+ \sigma_{\rm sn}^2} \, .
	\end{equation}
This reasoning can be generalized to other \carb nuclear spins of the environment by neglecting their mutual dipolar interaction with respect  to the hyperfine coupling with the NV center electronic spin. The Allan deviation expression then becomes 
	\begin{equation}
	\sigma_{\rm A}(\tau) = \sqrt{\sum_n\frac{{\mathcal{A}^{(n)}}^2}{4} (1-e^{-2\tau/T_R^{(n)}})+ \sigma_{\rm sn}^2} \, ,
	\label{eq_Allan_dev_instan_meas}
	\end{equation}
where $\mathcal{A}^{(n)}$ and $T_R^{(n)}$ refer to the hyperfine splitting and the relaxation time associated to the \carb nuclear spin $n$, respectively.

	\subsubsection{Allan deviation with averaging measurements}
		
In the previous section, we considered purely instantaneous measurements of the ESR frequency. However the time needed to record an individual ESR spectrum, $T_m \simeq 0.45$~s, can not be neglected in most experiments. This is especially true when $T_m$  is longer or in the same range as the correlation time of the nuclear spin environment. In this case, dynamics of the nuclear spin reservoir are partially erased by averaging over $T_m$. We now take into account the finite measurement time in the model of the Allan deviation. 

We start the analysis for a NV defect coupled with only one \carb nuclear spin and we consider that the ESR frequency measurement results from averaging over $M$ instantaneous measurements performed periodically every $\delta t$, such that $M \delta t = T_m$. The random variable $f_a(t)$ corresponds to the averaged frequency over $T_m$ which is expressed by 
	\begin{equation}
	f_a(t) = \frac{1}{M} \sum_{m=0}^{M-1} f(t+ m \delta t) = \frac{1}{M} \sum_{m=0}^{M-1} f(0) (-1)^{N(t+ m \delta t)} \, .
	\label{EqFa}
	\end{equation} 
We then express $N(t + m \delta t)$ in terms of independent random variables 
	\begin{equation}
	N(t+m \delta t) = N(t) + \sum_{k=1}^m N_k(\delta t) \, ,
 	\end{equation}
where $N_k(\delta t)$ denotes the number of nuclear spin flips occurring during the time interval $[t+ (k-1) \delta t, t + k \delta t]$. Using this transformation, Eq.~(\ref{EqFa}) can be written
	\begin{equation}
	f_a(t) = \frac{f(0)(-1)^{N(t)}}{M} \sum_{m=0}^{M-1} \prod_{k=1}^{m} (-1)^{N_k(\delta t)}\, .
	\end{equation}
	
	Considering that the number of nuclear spin flips follows a Poissonian probability distribution [Eq.~(\ref{Poisson})], the average value of $f_a(t)$ reads
	\begin{subequations}
	\begin{align}
	\langle f_a(t) \rangle & =  \frac{f(0)\langle (-1)^{N(t)} \rangle }{M} \sum_{m=0}^{M-1} \prod_{k=1}^m \langle (-1)^{N_k(\delta t)} \rangle \\
		& =  \frac{f(0)e^{-2 t/T_R}}{M} \sum_{m=0}^{M-1} e^{-2m \delta t / T_R}\\
		& =  \frac{f(0)e^{-2 t / T_R}}{M}\cfrac{1-e^{-2M \delta t /T_R}}{1-e^{-2\delta t/T_R}} \, .
	\end{align}
	\end{subequations}
In the limit where $\delta t \rightarrow 0$ with $M \delta t = T_m$, this last equation becomes 
	\begin{equation}
	\langle f_a(t) \rangle  = f(0) e^{-2 t/T_R} \left( \cfrac{1-e^{-2T_m/T_R}}{2T_m/T_R}\right) \, .
	\end{equation}
Considering the limit $T_m \rightarrow 0$, then $\langle f_a(t) \rangle \rightarrow f(0) e^{-2t/T_R}$, that corresponds to the expression previously obtained for the instantaneous measurement [Eq (\ref{eq_f_inst_mean})]. On the other hand, if $T_m\gg T_R$, then $\langle f_a(t) \rangle \rightarrow 0$ because frequency shifts average to zero due to rapid flips of the \carb nuclear spin.

Using the same methodology, we can calculate the value of $\langle (f_a(t))^2 \rangle $ : 
	\begin{equation}
	\langle f_a(t)^2 \rangle =  f(0)^{2}\times \frac{2 T_m /T_R + e^{-2T_m/T_R}-1}{2T_m / T_R} \, ,
	\label{TestEq}
	\end{equation}
as well as the correlation term defined for $\tau \geq T_m$
	\begin{equation}
	\langle f_a(t) f_a (t+ \tau) \rangle =  f(0)^{2}\times e^{-2 \tau/T_R} \left( \cfrac{\sinh (T_m/T_R)}{T_m/T_R} \right)^2 \, .
\label{TestEq2}
	\end{equation}
	
In the limit $T_m\ll T_R$, we get back to the case of purely instantaneous measurements $\langle f_a(t) f_a (t+ \tau) \rangle=f(0)^2 e^{-2 \tau /T_R}$ [see Eq.~(\ref{eq_f_corr})] and for $T_m\gg T_R$ we find  $\langle f_a(t) f_a (t+ \tau) \rangle \rightarrow 0$, as expected. In this limit, the correlation goes to zero as each of the measurements $\langle f_a(t)\rangle$ goes to zero.

Using Eq. (\ref{eq_Allan_dev_temp}), the Allan deviation can finally be expressed as : 
	\begin{equation}
	\sigma_{\rm A}(\tau)= \sqrt{\frac{{\mathcal{A}}^2}{4}  (\alpha-\beta^2 e^{-2(\tau-T_m)/T_R})+ \sigma_{\rm sn}^2} \, ,
	\label{eq_Allan_dev_fit}
	\end{equation}
	with 
	\begin{equation}
		\left\{
		\begin{aligned}
		\alpha & = \frac{T_R}{T_m} (1- \beta) \\ 
		\beta &= \frac{1- e^{-2 T_m/T_R}}{2 T_m/T_R}
		\end{aligned}
		\right. 
	\end{equation}
Extending the calculation to several independent \carb nuclear spins of the environment, this expression transforms to
	\begin{equation}
		\sigma_{\rm A}(\tau) = \sqrt{\sum_n\frac{{\mathcal{A}^{(n)}}^2}{4} (\alpha^{(n)}-{\beta^{(n)}}^2 e^{-2(\tau-T_m)/T_R^{(n)}})+ \sigma_{\rm sn}^2} \, ,
	\label{eq_Allan_dev_final}
	\end{equation}
	with
	\begin{equation}
		\left\{
		\begin{aligned}
		\alpha^{(n)} & = \frac{T_R^{(n)}}{T_m} (1- \beta^{(n)}) \\ 
		\beta^{(n)} &= \frac{1- e^{-2 T_m/T_R^{(n)}}}{2 T_m/T_R^{(n)}} \ .
		\end{aligned}
		\right. 
	\end{equation}

The hyperfine splitting of the ESR line is given by $\mathcal{A}^{(n)}=\sqrt{[\mathcal{A}_{ani}^{(n)}]^2+[{\mathcal{A}}_{zz}^{(n)}-\gamma_{n}B]^{2}}-\gamma_{n}B$, where $\mathcal{A}_{ani}^{(n)}$ and $\mathcal{A}_{zz}^{(n)}$ are the anisotropic and longitudinal components of the hyperfine tensor, respectively, and $\gamma_{n}\approx 1.07$~kHz/G is the $^{13}$C gyromagnetic ratio~[\onlinecite{Dreau2012}]. In the limit of high magnetic fields, {\it i.e.} for $\gamma_n B\gg(\mathcal{A}_{zz}^{(n)}, \mathcal{A}_{ani}^{(n)})$, the hyperfine splitting simplifies as $\mathcal{A}^{(n)}=\mathcal{A}_{zz}^{(n)}$.

Using $\tau = k \cdot T_m$ and replacing $\mathcal{A}^{(n)}$ by $\mathcal{A}_{zz}^{(n)}$ in Eq.~(\ref{eq_Allan_dev_final}), we finally obtain the expression $\sigma_{\rm A}(k)$ given in Eq. (3) of the main text. 

The limit of large $\tau$ gives the expected value for the plateau of the Allan deviation :
	\begin{equation}
\sigma_{\rm A}(\infty) = \sqrt{\sum_n\frac{{\mathcal{A}^{(n)}}^2}{4} \alpha^{(n)}+ \sigma_{\rm sn}^2} \, .
\label{AllanPlateau}
	\end{equation}

We note that he description of correlations in terms of Allan variance is equivalent to the description in terms of standard correlation function. Thus, the Allan variance is linked to the nuclear spin noise spectrum through a Fourier transform. 

\begin{figure*}[t]
    \begin{center}
        \includegraphics[width=0.88\textwidth]{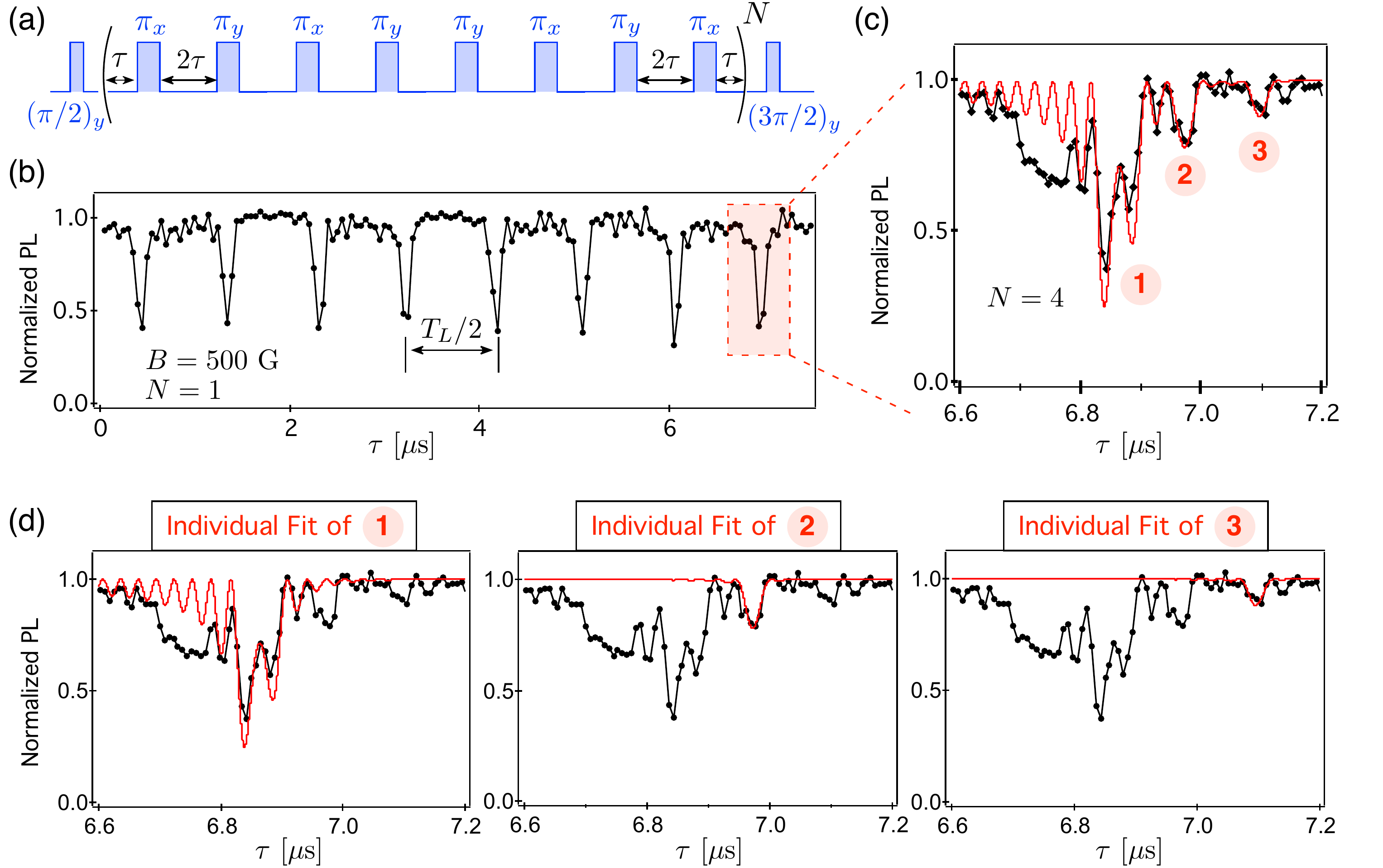}
    \end{center}
    \caption{(a) XY8 dynamical decoupling sequence. The duration of the microwave $\pi$-pulses is set to $\sim 100$~ns. The NV defect electronic spin is initially prepared in a coherent superposition with a $(\pi/2)_y$ pulse. At the end of the decoupling sequence, the coherence is probed by applying a $(3\pi/2)_y$ pulse. (b) Coherence signal as a function of $\tau$ recorded from NV2 for a decoupling sequence with $N=1$ and a magnetic field $B=500$~G applied along the NV defect quantization axis. The delay between two successive resonances is given by $T_{L}/2$, where $T_{L}$ is the Larmor precession period of the \carb nuclear spins. (c) Zoom in the $8^{\rm th}$ order resonance for $N=4$ units of the decoupling sequence. The red solid line is data fitting while considering three \carb nuclear spins coupled to the NV center. In (d), we represent the result of the fit for each individual \carb.}
    \label{fig:figureSI}
\end{figure*}	

	\subsection{Link between the Allan deviation and the ESR frequency histograms}
		
The histograms shown in Fig. 2(b) of the main manuscript indicate the distribution of the resonant frequencies extracted by fitting each individual ESR spectra with a Gaussian function. This section analyse the relationship between the profile of these histograms and the Allan deviations shown in Fig. 3(c).

By using Eq.~(\ref{test}) in the limit of large $\tau$, the Allan deviation is given by 
\begin{equation}
\sigma_{\rm A}(\infty) = \sqrt{\left \langle f(t)^2 \right \rangle} \, .
\end{equation}
The plateau of the Allan deviation is therefore equal to the standard deviation of the ESR frequency histogram.
When the measurement time $T_m$ is much larger than the correlation time of the bath, {\it i.e.} for $T_m\gg T_R$, the histogram of ESR frequencies is well described by a Normal distribution with a standard deviation set by the measurement noise $\sigma_{\rm sn}$. This case is obtained at low magnetic field ($B=300$~G) in Figure 2(b). Fitting the resulting histogram with a Normal distribution leads to a standard deviation $\sigma=\sigma_{\rm sn}=51\pm 1$~kHz. According to Eq.(~\ref{AllanPlateau}), the Allan deviation is expected to be flat in a regime without correlations, with a value given by $\sigma_{\rm A}=\sigma_{\rm sn}$. This is exactly what is observed experimentally [Fig. 3(c) of the main manuscript]. We note that $\sigma_{\rm sn}$ was then fixed to $51$~kHz for fitting Allan deviation measurement recorded at high field in Fig. 3(c). 

To understand the profile of the histograms at higher magnetic fields and their link with the Allan deviation, we first consider a single \carb nuclear spin of the reservoir with a relaxation time $T_R$ such that $T_m \ll T_R$. In that case, the histogram of the ESR frequencies results from the sum of two Normal distributions centered at $\pm \mathcal{A}/2$. If $\mathcal{A}>\sigma_{sn}$, these distributions can be well resolved leading to two peaks in the histogram, as observed for NV1 in Fig. 2(b) at high magnetic fields. By considering only a single \carb of the bath, each peak would be described by a Normal distribution with standard deviation $\sigma_{sn}$. However, multiple nuclear spins from the reservoir are coupled with the NV defect with hyperfine coupling strengths $\mathcal{A}^{(n)}$ and relaxation times $T_R^{(n)}$. When the magnetic field strength is such that the nuclear Zeeman term becomes much larger than the anisotropic hyperfine component, the relaxation time $T_R^{(n)}$ can exceed the measurement time, leading to a more complex distribution. If $\mathcal{A}^{(n)}<\sigma_{\rm sn}$, the resulting splitting can not be resolved in the histogram and correlations between measurements are evidenced by an overall broadening of each peaks of the histogram. This is experimentally observed in Figure 2(b) when the magnetic field is increased, which indicates a contribution to the Overhauser field fluctuations from multiple \carb nuclear spins.  We note that in this case each peak of the histogram can not be described by a simple Gaussian function because it results from the sum of Normal distributions. The Allan deviation plateau remains equal to the standard deviation of the full ESR frequency distribution, which depends on the hyperfine strengths and the correlation times following Eq.(~\ref{AllanPlateau}). 

\subsection{Dynamical decoupling signal}


The hyperfine coupling strength between the NV defect electronic spin and nearby \carb of the reservoir can be precisely characterized by using dynamical decoupling methods~\cite{Taminiau2012,Kolkowitz2012,Zhao2012}. Here we apply $N$ units of the symmetric XY8 pulse sequence [Fig.~\ref{fig:figureSI}(a)] to the single NV defect denoted NV2 in the main text. A typical experimental signal obtained for $N=1$ is depicted in Fig.~\ref{fig:figureSI}(b), revealing the characteristic collapses of the coherence signal at the Larmor frequency. By zooming inside the $8^{\rm th}$ collapse while applying a decoupling sequence with $N=4$ units, sharp dips can be clearly distinguished [Fig.~\ref{fig:figureSI}(c)]. These dips correspond to resonances with individual \carb precessing at different speeds. The position and the amplitude of each dip are directly linked to the longitudinal and anisotropic components of the hyperfine interaction for a particular \carb. The experimental data were then fitted by following the procedure described in Ref.~(\onlinecite{Taminiau2012}). The values of the hyperfine coupling extracted from these fits are summarized in table I of the main text for the three strongest nuclear spins coupled to the NV center. We find that the $^{13}$C labeled number 1 has a strong anisotropic hyperfine coupling ($\mathcal{A}_{ani}^{(1)}=128\pm2$~kHz) that leads to oscillations on the dynamical decoupling signal. On the other hand, $^{13}$C labeled number 2 and 3 have a much weaker anisotropic interaction, leading to single resonance peaks [Fig.~\ref{fig:figureSI}(d)].

\begin{figure}[h!]
    \begin{center}
        \includegraphics[width=0.5\textwidth]{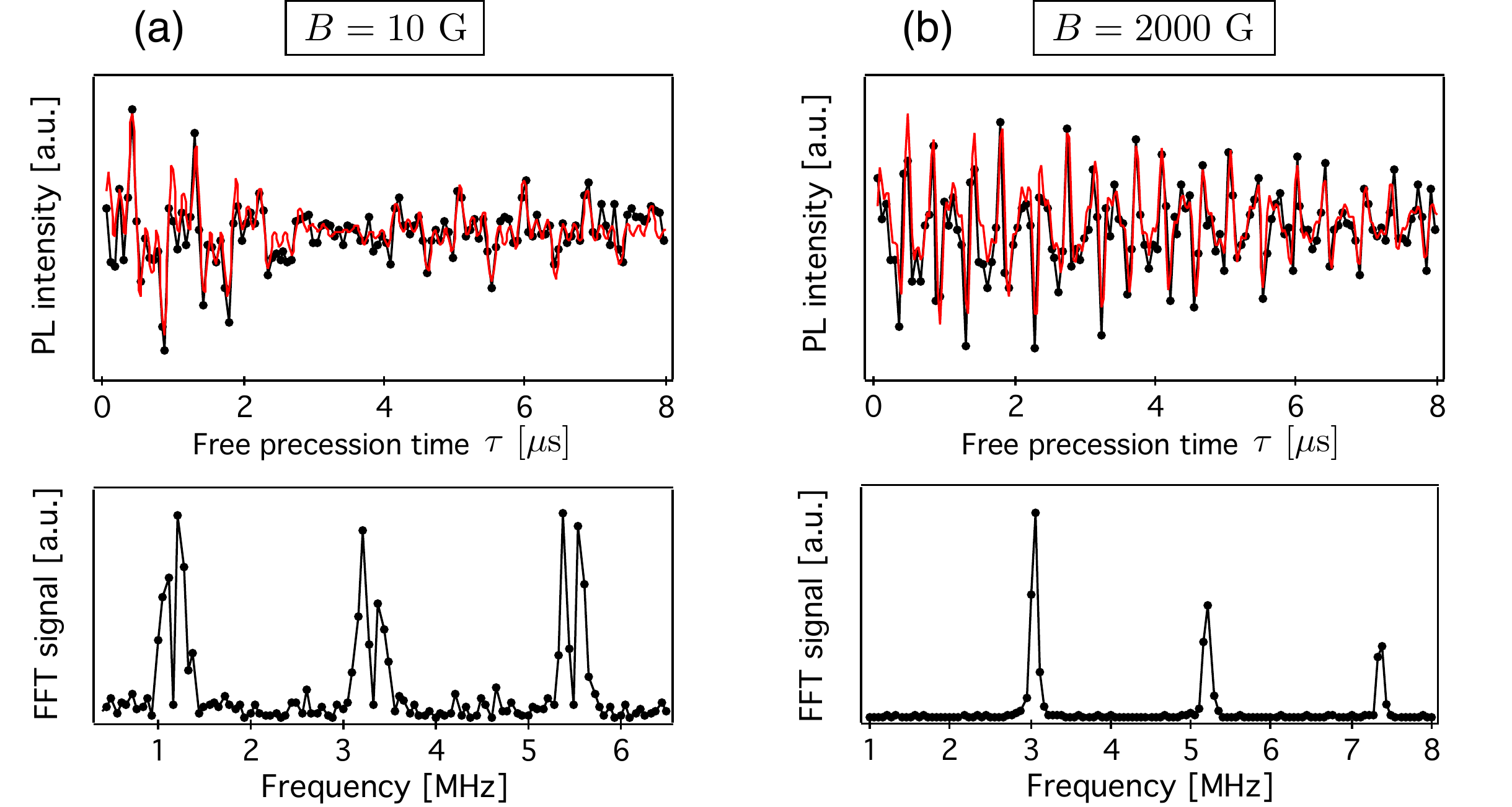}
    \end{center}
    \caption{Ramsey fringes and corresponding Fourier transform spectrum recorded for NV2 at (a) $B=10$~G and (b) $B=2000$~G.}
    \label{fig:figureSI2}
\end{figure}

The strong value of the anisotropic component measured for $^{13}$C number 1 was further confirmed through Ramsey measurements, by using the usual sequence consisting in two microwave $\pi/2$-pulses separated by a variable free evolution duration $\tau$. Here we try to detect directly the hyperfine splitting induced by this particular \carb. As indicated above, this splitting is given by $\mathcal{A}=\sqrt{[\mathcal{A}_{ani}^{(1)}]^2+[{\mathcal{A}}_{zz}^{(1)}-\gamma_{n}B]^{2}}+\gamma_{n}B$. At low field, $\mathcal{A}\approx \mathcal{A}_{ani}^{(1)}$ since ${\mathcal{A}}_{zz}^{(1)}\ll\mathcal{A}_{ani}^{(1)}$ [See Tab.~(1) of the main text]. Ramsey fringes together with their Fourier transform spectrum recorded at $B=10$~G are shown in Fig.~\ref{fig:figureSI2}(a). Apart from the usual hyperfine structure linked to the $^{14}$N nuclear spin of the NV defect, we observe an additional hyperfine splitting $\mathcal{A}\sim 150$~kHz, as expected. In the limit of high magnetic fields, the hyperfine splitting becomes $\mathcal{A}=\mathcal{A}_{zz}^{(1)}$. Since the longitudinal hyperfine component is weak for $^{13}$C number 1, its hyperfine splitting can not be observed in the FFT spectrum of Ramsey fringes recorded at $B=2000$~G [Fig.~\ref{fig:figureSI2}(b)].

\end{document}